\documentclass[pdflatex,sn-mathphys-num]{sn-jnl}

\usepackage{graphicx}%
\usepackage{multirow}%
\usepackage{amsmath,amssymb,amsfonts}%
\usepackage{amsthm}%
\usepackage{mathrsfs}%
\usepackage[title]{appendix}%
\usepackage{textcomp}%
\usepackage{manyfoot}%
\usepackage{booktabs}%
\usepackage{algpseudocode}%
\usepackage{listings}%
\usepackage{mathrsfs}
\usepackage{array}
\usepackage[table]{xcolor}
\usepackage{colortbl}
\usepackage{tabularx}
\usepackage{array}
\usepackage{makecell}
\usepackage{soul}
\definecolor{meugraybox}{HTML}{EAEAEA}

\newsavebox\foobox
\newlength{\foodim}
\newcommand{\slantbox}[2][0]{\mbox{%
        \sbox{\foobox}{#2}%
        \foodim=#1\wd\foobox
        \hskip \wd\foobox
        \hskip -0.5\foodim
        \pdfsave
        \pdfsetmatrix{1 0 #1 1}%
        \llap{\usebox{\foobox}}%
        \pdfrestore
        \hskip 0.5\foodim
}}
\def\Laplace{\slantbox[-.45]{$\mathscr{L}$}}
\definecolor{BqnRow}{RGB}{228,236,255}
\definecolor{BHRow}{RGB}{255,204,204}

\theoremstyle{thmstyleone}%
\theoremstyle{thmstyletwo}%

\theoremstyle{thmstylethree}%

\begin{document}

\title[Article Title]{A hierarchical statistical mechanics approach: hyperstatistics and its applications}


\author[1]
{\fnm{Lucas} \sur{Squillante}}

\author[1]
{\fnm{Samuel M.} \sur{Soares}}

\author[2]
{\fnm{Guilherme} \sur{Lepski}}


\author*[1]{\fnm{Mariano} \sur{de Souza}}\email{mariano.souza@unesp.br}

\affil[1]
{\orgdiv{IGCE -- Physics Department}, \orgname{S\~ao Paulo State University (Unesp)}, \orgaddress{\city{Rio Claro}, \state{SP}, \country{Brazil}}}

\affil[2]{\orgdiv{Laboratory of Experimental Surgery (LIM26)}, \orgname{Medical School, University of S\~ao Paulo}, \orgaddress{\city{S\~ao Paulo}, \country{Brazil}}}


\abstract{In a recent paper [arXiv:2604.24783v5\,(2026)], we have proposed a general approach to treat systems with inherent non-Boltzmann-Gibbsian behaviour at the domain level.  Given the high accuracy of our approach and based on the fact that the generalized Boltzmann factor at the macroscopic level embodies a  hypergeometric function, we have adopted the term hyperstatistics. We have applied such a statistical mechanics approach to the discharge of a capacitor in a RC series circuit, pressure decay over time during the pumping of $^4$He lines of a closed cycle cryostat, transverse momentum distribution in midrapidity data of $p$-Pb collisions at the LHC, as well as for the distribution of accelerations in turbulent systems. Here, we discuss into more details the ground of hyperstatistics. We demonstrate the versatility of hyperstatistics upon applying it to the velocity autocorrelation function in Brownian motion and also regarding its potential to describe brain dynamics.}




\maketitle

\section{Introduction to hyperstatistics and a few remarks}\label{introduction}
\begin{figure}[!b]
\centering
\includegraphics[width=0.6\textwidth]{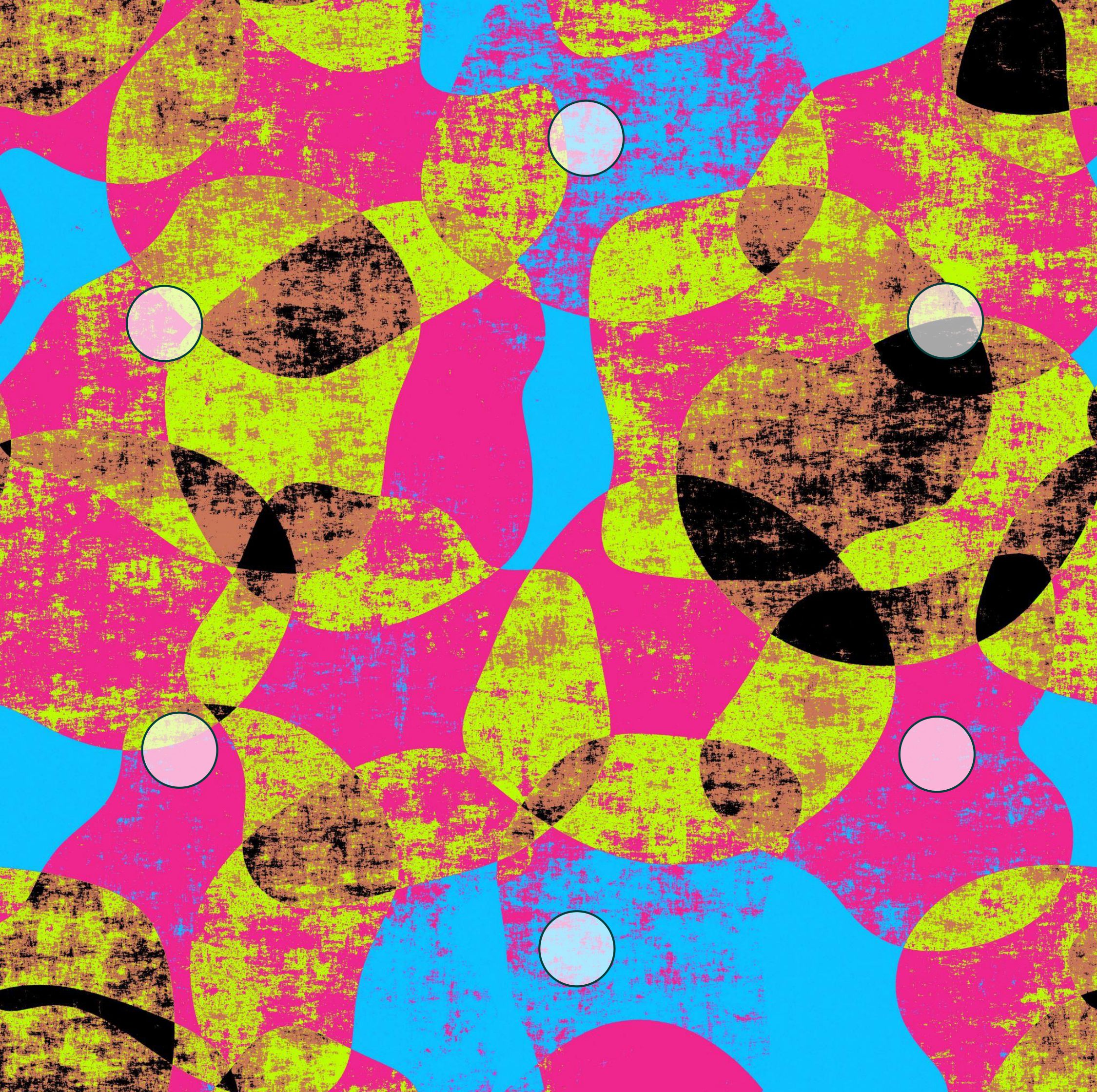}
\caption{\footnotesize Hypothetical representation of a complex system: different colours represent, for instance, distinct temperatures with their corresponding Boltzmann factors, which can be associated to each pixel. The six circles pinpointed on distinct regions represent a device to measure the observable of interest. Each measurement performed in a given region of the system mirrors the average of such observable in the corresponding domain. This simple scheme illustrates that upon carrying out a measurement in a real system, we are accessing the expected value of the quantity of interest of a given domain, cf.\,Eq.\,14 of Ref.\,\cite{arxivhyper}.}
\label{Fig-1}
\end{figure}

The term temperature is broadly used in our daily life. Examples of such usage include weather forecast and checking whether a person has a fever. 
It turns out that upon measuring the temperature of a system, whether using a thermometer attached directly to the sample, as usual in solid state Physics, or positioned in its vicinity \cite{reviewmariano}, the recorded temperature does not correspond to the one of a specific point, but rather to a comprising region of the system of interest. As mentioned in Ref.\,\cite{valvano1992} by Valvano, under a biological perspective, it reads ``... \emph{Another way to define spatial sensitivity is to consider a probe which is placed in a tissue having spatial temperature gradients. The probe output will be some average of the surrounding temperature.} ...''. In some cryostats, for instance, the thermometer is placed close to the sample and it is generally assumed that the sample temperature is the same as in its vicinity, which in reality might not be the case. Hence, as previously mentioned, temperature measurements might be seen as an average of the temperatures in the surrounding area of the point of interest \cite{valvano1992}, cf.\,Fig.\,\ref{Fig-1}, being to some extent an analogous situation found in experiments employing scanning tunnelling spectroscopy to access the electronic density of states \cite{hitosugi2016}. Also, given the inherent competition of phases on the verge of phase transitions, a distribution of temperatures has been considered in this regime \cite{prl07,GR1}.  
This is why, under an experimental perspective, in the light of hyperstatistics, a distribution of Boltzmann factors within a given domain of the system gives rise to a $q$-exponential ($q$-exp) function expressed in terms of an average inverse temperature $\langle \beta_i \rangle$ \cite{arxivhyper}, where $\beta_i$ denotes the inverse temperature of an $i$-th domain. This is because when hypothetically  measuring the temperature at a particular point, e.g., at the center of a given domain, the measured value of temperature corresponds to an average over the slightly different temperatures in its surroundings, yielding the value $\langle \beta_i \rangle$. In the frame of hyperstatistics, the ($q$, $n$)-generalized effective Boltzmann factor at the domain level $B_q^{n}(E) = \int_{0}^{\infty}\exp(-\beta_i E) f(\beta_i)d\beta_i$ simply results in $B_q^{n}(E) = \exp_{q}(-\langle \beta_{i} \rangle E)$ for the $\gamma$-distribution \cite{arxivhyper}, where $E$ is the energy. This is why we have stated in Ref.\,\cite{arxivhyper} that $\langle \beta \rangle = \langle \langle \beta_i \rangle \rangle$ actually corresponds to the average inverse temperature of the whole system. In the mesoscopic (domain) level, as previously mentioned, the value $\langle \beta_i \rangle$ represents the inverse temperature of a given domain, so that $\langle \beta \rangle$ in turn corresponds to the average taken over all $\langle \beta_i \rangle$ in the macroscopic scale, cf.\,Table\,\ref{Table-1}. However, when measuring the temperature of a system, we only have access to a local temperature value rather than the temperature of the whole system, cf.\,Fig.\,\ref{Fig-1}. This is why in hyperstatistics $B_q^n(E)$ is given in terms of $\langle \beta_i \rangle$ and not $\langle \beta \rangle$ \cite{arxivhyper}. Note that $\langle \beta_i \rangle$ explicitly depends on the chosen $f(\beta_i)$ \cite{arxivhyper}.
Essentially, in hyperstatistics we are dealing with three distinct size scales to treat the system of interest: sub-domains (microscopic), domains (mesoscopic), and the whole (macroscopic) system, cf.\,Table\,\ref{Table-1}. 
\begin{table*}[!b]
\centering
\begin{tabular}{@{}c@{}}
\includegraphics[width=0.9\textwidth]{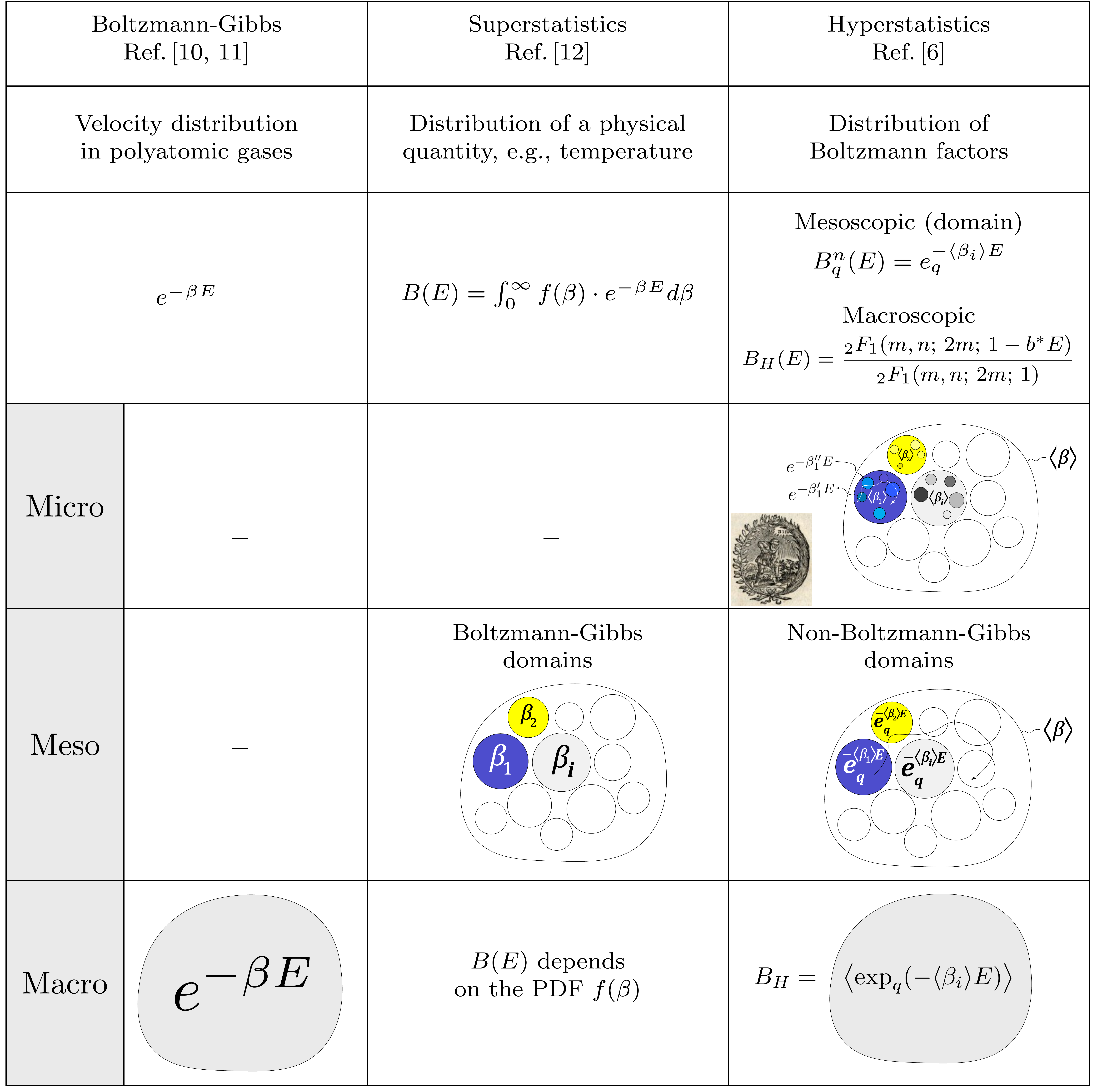}
\end{tabular}
\caption{\footnotesize \textbf{Foundations of distinct statistical mechanics approaches}. The ground of Boltzmann-Gibbs \cite{boltzmann,gibbs}, superstatistics \cite{beck2003}, and hyperstatistics \cite{arxivhyper} is depicted in each corresponding column. Schematic representations of the different characteristic size scales associated with each approach are presented. In the column corresponding to hyperstatistics, we include the original illustration from the cover of Ren\'e Descartes' book ``\emph{Discours de la m\'ethode}'' \cite{descartes}, in which a man is depicted digging ever deeper into the ground in search of the hidden truth, in direct analogy with the hierarchical levels incorporated in hyperstatistics.}
\label{Table-1}
\end{table*} 
Essentially, a $\gamma$-distribution \cite{wilk2000} of the canonical exponential weight associated with a corresponding physical quantity of interest within a single domain is considered. It turns out that upon considering an individual pixel depicted in Fig.\,\ref{Fig-1}, it can be described by a single canonical exponential weight, but this is not necessarily the case for the mesoscopic and macroscopic levels as it occurs in complex systems. In the case of a distribution of temperatures, the exponential weight is the Boltzmann factor. Hence, considering the temperature as the physical quantity of interest, such a protocol leads naturally to a $q$-exp function-based Boltzmann factor for the corresponding domain, including the emergence of $\langle \beta_i \rangle$ in its argument. A relevant mathematical aspect is that the $\gamma$-based Boltzmann factor distribution within a given domain represents the Laplace transform of the $\gamma$-distribution, being the Boltzmann factor its kernel \cite{arxivhyper}. At this point, it is worth recalling the canonical Laplace transform \cite{ricieribook}:
\begin{equation}\label{Laplacetransform}
\Laplace\{\ \!\!f(x)\} = \int_{0}^{\infty}f(x)e^{-xs} dx,
\end{equation}
where $f(x)$ is the mathematical function to be Laplace transformed, which in turn should ensure the integrability of Eq.\,\ref{Laplacetransform}. In hyperstatistics, we consider at the domain level:
\begin{equation}\label{Laplacehyper}
\Laplace\{\ \!\!\gamma(\beta_i)\} = \int_{0}^{\infty}\gamma(\beta_i) e^{-\beta_i E} d\beta_i = \int_{0}^{\infty} \left[\frac{1}{\Gamma(n)}
\left(\frac{n}{\langle \beta_i \rangle}\right)^n
\beta_i^{n-1}e^{-\frac{n\beta_i}{\langle \beta_i \rangle}}\right] e^{-\beta_i E} d\beta_i = e_q^{-\langle \beta_i \rangle E},
\end{equation}
where $e_q^{-\langle \beta_i \rangle E} = [1+(q-1)\langle \beta_i \rangle E]^{1/(1-q)}$ is a $q$-exp function, which recovers a classical exponential for $q \rightarrow 1$ \cite{tsallis}, i.e., $\lim_{q\rightarrow1}e_q^{-\langle \beta_i \rangle E} = e^{-\langle \beta_i \rangle E}$. Note that Eq.\,\ref{Laplacehyper} can be modified accordingly depending on the distributed quantity, so that considering a distribution of the inverse of the relaxation time $\tau$ and corresponding exponential weights, we have \cite{arxivhyper}:
\begin{equation}
\Laplace\{\gamma(1/\tau)\} = \int_0^{\infty}  e^{-t/\tau} \gamma(1/\tau) d(1/\tau) = e_q^{-t/\langle \tau \rangle},
\label{laplace2}
\end{equation}
where $t$ is the time. Equation\,\ref{laplace2} demonstrates that the non-Boltzmann-Gibbsian character of the domain can be described in terms of a ($q$, $n$)-generalized Boltzmann-type factor $B_q^n(t)$ that depends on the average relaxation time $\langle \tau \rangle$, which in turn also incorporates both $q$ and the degree $n$ of the $\Gamma$-function as independent parameters, namely \cite{arxivhyper}:
\begin{equation}
B_q^n(t) = \exp_q[- t / \langle \tau \rangle] = \exp_q\left\{-\frac{(n-1)}{(n\tau_0)[1+n(1-q)]} t\right\},
\label{BqE}
\end{equation}
where $\tau_0$ is a characteristic relaxation time, which is obtained considering the value of the physical observable of interest at 37\,\% of its initial value, analogously to the voltage decay in a capacitor \cite{arxivhyper}. The expression for $\langle \tau \rangle$ was obtained upon considering a $q$-$\gamma$-distribution of $1/\tau$ \cite{arxivhyper}. Note that, cf.\,discussions in Ref.\,\cite{arxivhyper}, $B_q^n$ should be employed for quantities measured locally, such as temperature, while for quantities experimentally determined in a macroscopic way, e.g., pressure or voltage decay in a capacitor, the following expression of the hyperstatistical generalized Boltzmann factor for the macroscopic level $B_H$ should be employed \cite{arxivhyper}:
\begin{equation}
B_H(E) = \frac{{}_2F_1(m, n; 2m; 1 - b^*E)}{{}_2F_1(m, n; 2m; 1)},
\end{equation}
where $_{2}F_1$ is a hypergeometric function, $m = 1/(q-1)$, and $b^* = \langle \beta \rangle/n$. It is worth mentioning that hyperstatistics has already been applied to other physical situations than those reported in Ref.\,\cite{arxivhyper}, e.g., the Klein-Gordon and Dirac oscillators \cite{boumali2026}, non-Markovian quantum decays \cite{nicola}, and collisionless-plasma Landau damping \cite{aboumalilandau}.

Another key aspect worth highlighting regarding the Laplace transform of the $\gamma$-distribution is that it embodies a ``competition'' between a power-law in terms of $n$ and an exponential decay, cf.\,the integrand of Eq.\,\ref{Laplacehyper}. Also, the $\gamma$-distribution is special because it incorporates a sum of exponential weights associated with the Poisson distribution \cite{arxivhyper}, which is a key ingredient embedded in the proposal of hyperstatistics. As previously discussed, the result of the integral in Eq.\,\ref{Laplacehyper} naturally leads to a $q$-exp function, which in turn incorporates $q$ and $n$ \cite{arxivhyper}. Hence, the values of $q$ and $n$ determine whether the analyzed physical decay is mainly dictated by a power-law or by a $q$-exp function. Such a ``competition'' can also be seen in the Mellin transform of the $q$-exp function \cite{arxivhyper}, which we have recognized as $\Gamma(n,q)$ \cite{arxivhyper}.
It is clear that the mathematical structure of Eq.\,\ref{Laplacehyper} is the Laplace transform of the $\gamma$ probability density function and the Boltzmann factor the kernel, cf.\,Eq.\,\ref{Laplacetransform}.
\begin{table}[b]
\centering
\renewcommand{\arraystretch}{1.2}
\setlength{\tabcolsep}{\dimexpr(\linewidth-10cm-5\arrayrulewidth)/8\relax}
\begin{tabular}{|>{\centering\arraybackslash}m{2.6cm}
                |>{\centering\arraybackslash}m{4.2cm}
                |>{\centering\arraybackslash}m{1.6cm}
                |>{\centering\arraybackslash}m{1.6cm}|}
\hline
\textbf{Experiment} & \textbf{Physical observable} & $q =$ & $n =$ \\
\hline
\rowcolor{BqnRow}
$p$-Pb\par collisions\par Ref.~\cite{alice}& Midrapidity & 1.141 & 4.553 \\
\hline
\rowcolor{BqnRow}
Turbulent\par systems\par Ref.~\cite{bodenschatz}& Acceleration & 1.179 & 2.908 \\
\hline
\rowcolor{BHRow}
Discharge of a\par capacitor\par Ref.~\cite{arxivhyper}& Voltage & 1.050 & 4.032 \\
\hline
\rowcolor{BHRow}
Pressure\par decay\par Ref.~\cite{arxivhyper}& Pressure & 1.097 & 2.842 \\
\hline
\rowcolor{BqnRow}
Brownian\par motion\par Ref.~\cite{brownian1}&
Parallel (perpendicular) component of the VACF & 1.613 (1.389) & 1.327 (1.693) \\
\hline
\end{tabular}
\caption{\textbf{Hyperstatistics for distinct experiments}. Relationship between the physical observables associated with each experiment \cite{arxivhyper,bodenschatz,alice,brownian1} and the corresponding values of $q$ and $n$. The experimental data sets employed to obtain $q$ and $n$ were reported in the indicated references. Rows shaded in blue (red) refer to fits performed with $B_q^n$ ($B_H$), cf.\,Fig.\,\ref{Fig-1.1}.}
\label{Table-2}
\end{table}
\begin{figure}[t!]
\centering
\includegraphics[width=0.8\textwidth]{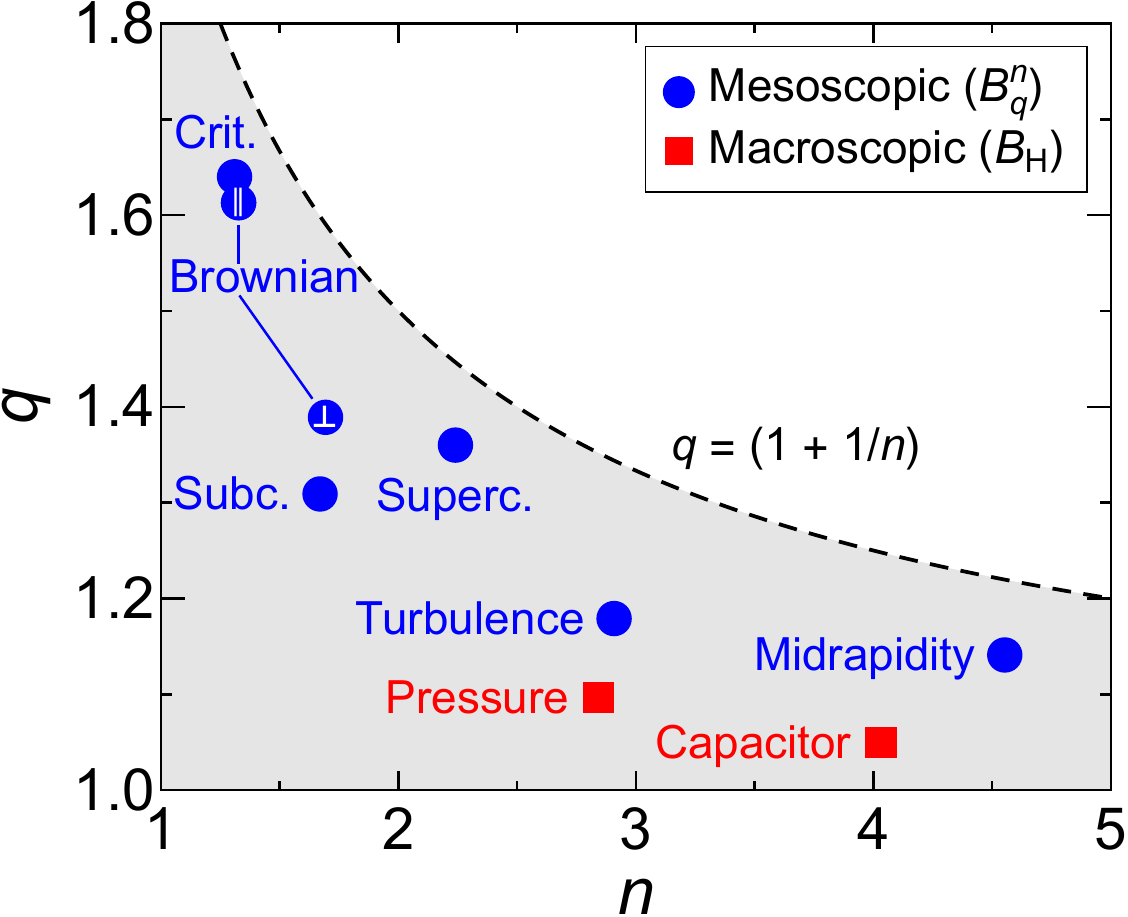}
\caption{\footnotesize Values of $q$ \emph{versus} $n$ obtained from the various fits depicted in Table\,\ref{Table-2} \cite{arxivhyper} and Section\,\ref{cbh}, where Subc., Crit., and Superc. refer to the subcritical, critical, and supercritical regimes, respectively. It is also indicated whether each pair $(q,n)$ was obtained using $B_q^n$ or $B_H$, cf.\,indicated in the label. The dashed line represents the boundary $q=(1+1/n)$ \cite{youtubevideo}, while the shaded gray region indicates the convergence regime in which hyperstatistics is applicable.}
\label{Fig-1.1}
\end{figure} 
Such a Laplace transform corrects the Boltzmann factor itself, enabling us to transit from a Boltzmann-Gibbsian to a hyperstatistics space, which embodies $q$ and $n$. This is quite remarkable because the Laplace transform usually only changes the variable space \cite{arxivhyper}. However, in this case, the kernel itself, namely the Boltzmann factor, is corrected, resulting in a $q$-exp function in terms of $\langle \beta_{i} \rangle$, cf.\,Table\,\ref{Table-1}. Yet, analogously to the vision of a rhino being corrected as a result of a Laplace transform \cite{ricieribook}, the $q$-exp function is the corrected Boltzmann factor at the domain level. As discussed in Ref.\,\cite{arxivhyper} regarding the proposal of hyperstatistics, the $q$-generalized $\Gamma$ function, i.e., $\Gamma(n, q)$, implies the convergence criterion $1 < q < (1 + 1/n)$, cf.\,Fig.\,\ref{Fig-1.1}. The value of $n$ determines the upper bound of $q$, making it evident that $q$ and $n$ are related quantities through such an inequality. In other words, a given value of $n$ implies a corresponding range of allowed values of $q$. For instance, in Fig.\,4 of Ref.\,\cite{arxivhyper}, regarding the midrapidity data set, the values of $n = 4.553$ and $q = 1.141$ were obtained, which lead to $1 < q < 1.2196$, in agreement with the obtained value of $q$ for this particular case. Of course, when performing the Laplace transform of the $\gamma$-distribution, the equality $n = 1/(q-1)$ must be employed in order to obtain a $q$-exp function \cite{arxivhyper}. However, this equality refers only to the upper boundary condition and may therefore be interpreted as a limiting case \cite{youtubevideo}, cf.\,Fig.\,\ref{Fig-1.1}. Note that the expected value of the physical quantity of interest embedded in the $q$-exp function in Eq.\,\ref{BqE}, namely $\langle \tau \rangle = (n\tau_0)[1+n(1-q)]/(n-1)$, only recovers $\tau_0$ for $q \to 1$ and $n \to \infty$, i.e., $\lim_{n\to\infty}\lim_{q\to1} \langle \tau \rangle = \tau_0$. Note that the sequence of the application of the limit is relevant. This is the reason why $n$ and $q$ can be treated as independent fitting parameters when analyzing a particular data set. The independence of $q$ and $n$ emerges in the computation of the expected value of the physical quantity of interest. In the present work, we restrict our analysis to the mesoscopic (domain) level, i.e., all new data sets discussed here are fitted employing $B_q^n$ (Eq.\,\ref{BqE}). This choice follows the protocol established in Ref.\,\cite{arxivhyper}, namely that $B_q^n$ is used for quantities accessed locally, whereas $B_H$ is suitable for quantities measured over the system as a whole, such as the voltage decay of a capacitor or the pressure decay upon pumping a container. The velocity autocorrelation function discussed in Section\,\,\ref{VACFS} refers to a single optically trapped particle, so that it probes the dynamics of the particle and its immediate fluid surroundings, i.e., at the domain level, rather than an average over the whole suspension. Analogously, the avalanche-size distributions discussed in Section\,\ref{cbh} are treated as a local measurement, in direct analogy with electroencephalogram data set, in which each electrode accesses only the local activity of the brain. For the sake of completeness, the $(q, n)$ pairs obtained in Ref.\,\cite{arxivhyper} using $B_H$ for the capacitor discharge and the pressure decay are also included in Table\,\ref{Table-2} and Fig.\,\ref{Fig-1.1}. Next, we discuss in details how $q$ and $n$ affect $B_q^n$.

\section{The role played by $q$ and $n$ in $B_q^n(t)$}\label{qandn}
\begin{figure}[b!]
\centering
\includegraphics[width=0.8\textwidth]{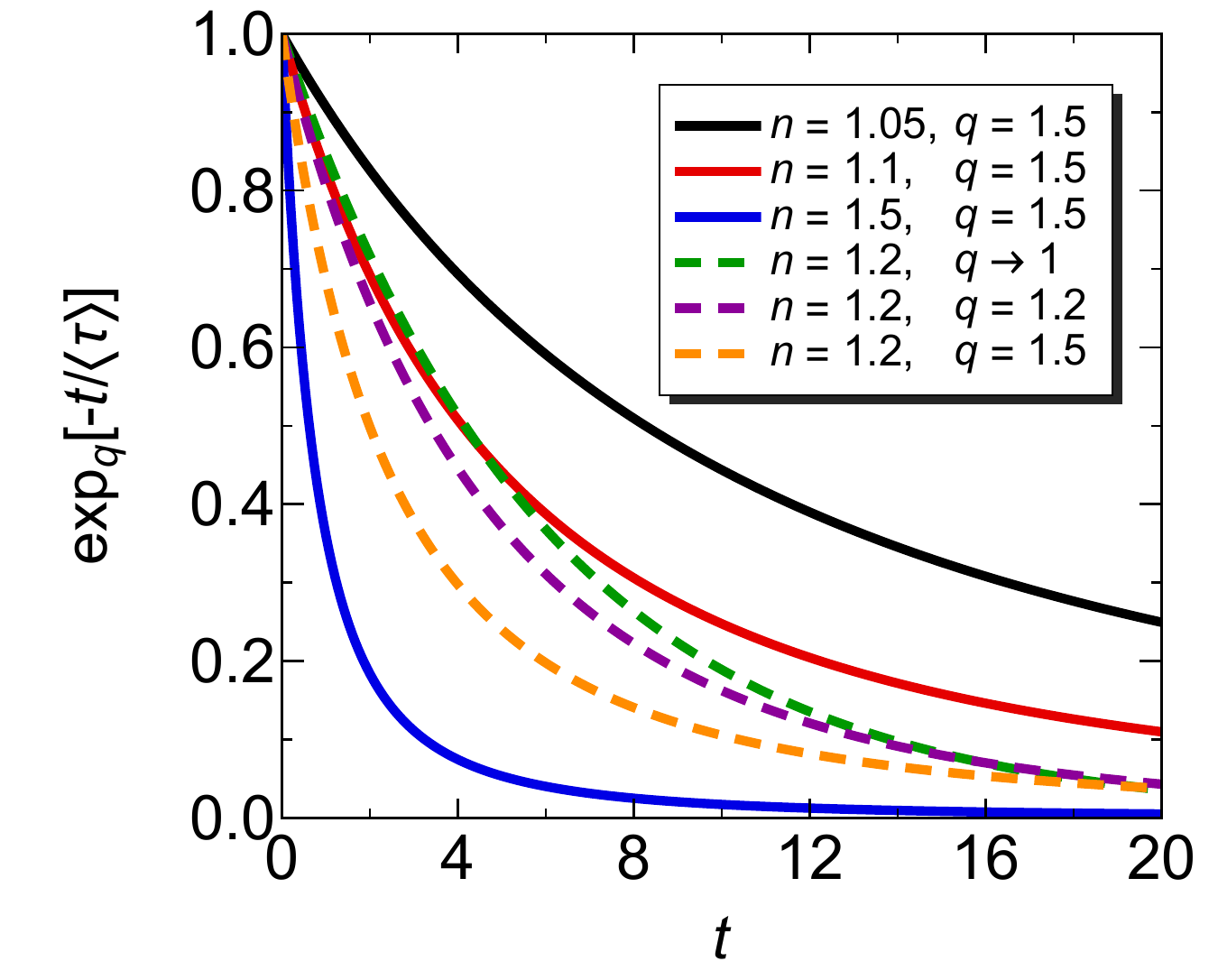}
\caption{\footnotesize $B_q^n(t) = \exp_q[-t / \langle \tau \rangle]$ \emph{versus} $t$ for $q = 1.5$ and $n = 1.05$ (solid black line), $n = 1.1	$ (red solid line), and $n = 1.5$ (blue solid line), as well as for a fixed value of $n = 1.2$ and $q \to 1$ (green dashed line), $q = 1.2$ (purple dashed line) and $q = 1.5$ (orange color dashed line). Note that $\exp_q[-t/\langle \tau \rangle] \rightarrow 1$ for $t \rightarrow 0$ regardless of $q$ and $n$. All curves were obtained following Eq.\,\ref{BqE}. More details in the main text.}
\label{Fig-2}
\end{figure}

In the following we analyse $B_q^n$ considering a $\gamma$-type distribution of $1/\tau$. As depicted in Fig.\,\ref{Fig-2}, for a fixed value of $q$, varying $n$ primarily affects the initial decay of $B_q^n(t)$. Note that upon decreasing $n$ leads to a slower decay, i.e., longer tail, of $B_q^n(t)$. 
\begin{figure}[t!]
\centering
\includegraphics[width=\textwidth]{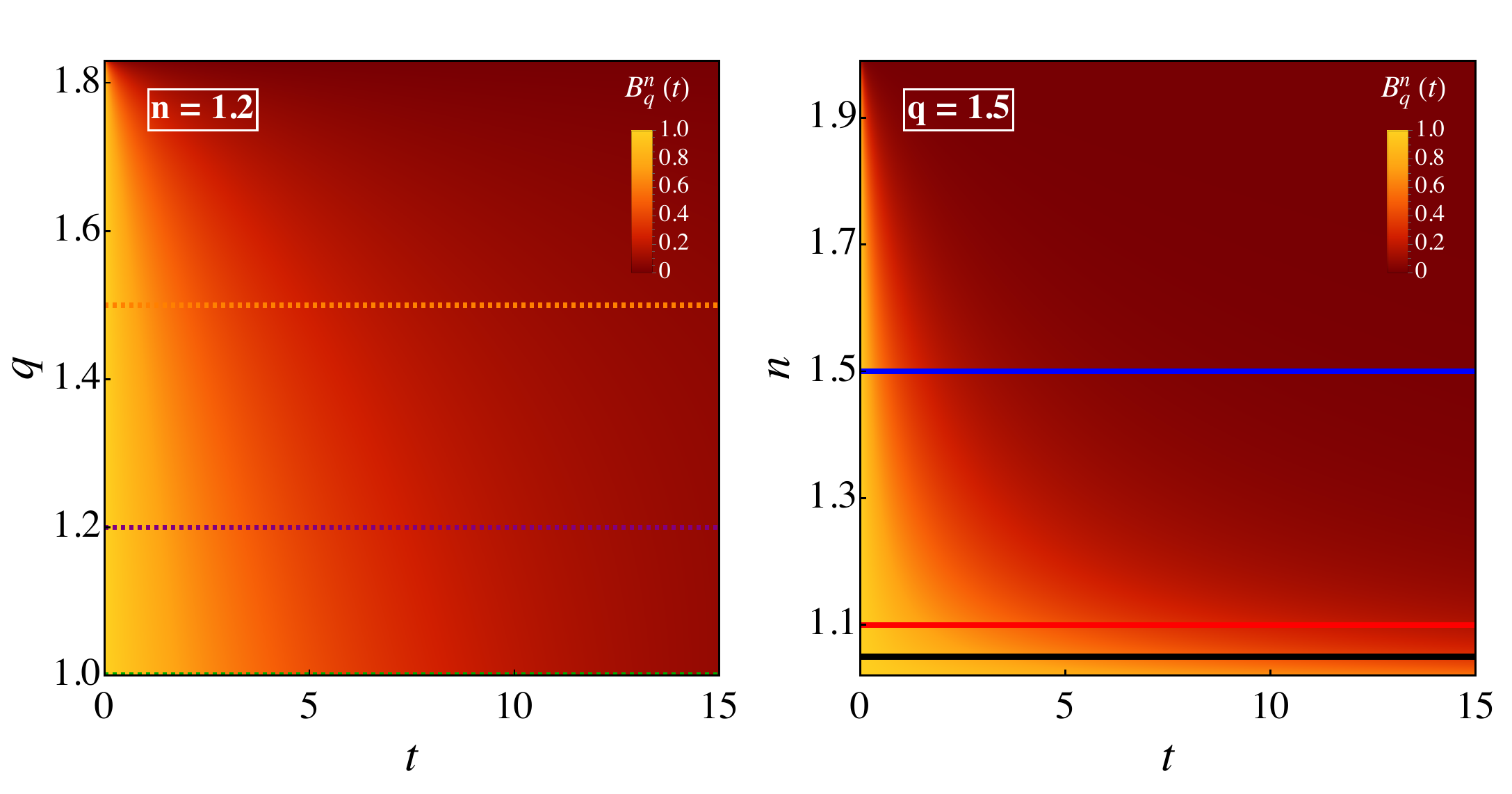}
\caption{\footnotesize Left panel: Density plot of $B_q^n(t)$ \emph{versus} $t$ \emph{versus} $q$ for a fixed value of $n = 1.2$. The dashed lines are shown in conformity with Fig.\,\ref{Fig-2}. Right panel: Density plot of $B_q^n(t)$ \emph{versus} $t$ \emph{versus} $n$ for a fixed value of $q = 1.5$ and the solid lines are also depicted in connection with Fig.\,\ref{Fig-2}. More details in the main text.}
\label{Fig-3}
\end{figure}
Note that the results shown in both Figs.\,\ref{Fig-2} and \ref{Fig-3} are in line with $n > 1$ and the convergence criterion $1 < q < [1 + (1/n)]$ of the $\Gamma(n,q)$ function \cite{arxivhyper}. This is because both $q$ and $n$ are embedded in the expected value of the physical quantity of interest. For a fixed value of $n$, as shown in Fig.\,\ref{Fig-2}, the parameter $q$ becomes important in the long-tail regime. The density plots shown in Fig.\,\ref{Fig-3} further illustrate this aspect in a more general way for a set of $n$ and $q$ values. Hence, $n$ and $q$ play complementary roles, i.e., while $n$ quantifies the degree of heterogeneity, $q$ predominantly controls the asymptotic behaviour of $B_q^n(t)$.  Hyperstatistics thus provides a versatile framework for describing decay processes in complex systems where non-Boltzmann-Gibbsian statistics are present at the domain level. 

Based on our discussions in Ref.\,\cite{arxivhyper}, the hyperstatistics approach can be applied to a wide range of physical situations, cf.\,Table\,\ref{Table-2}.  It is worth mentioning that in a previous work, we have considered a multilevel model to describe double-maximum anomalies in the specific heat, being each energy level associated with a distinct Boltzmann factor \cite{BJP}. Such a simple model has been successfully employed in the specific heat in Ce-based systems, see, e.g., Ref.\,\cite{Jin}. However, a more robust approach to treat complex systems is required and thus hyperstatistics comes into play. Regarding the field of mathematical biosciences, hyperstatistics can be employed to describe biomolecular diffusional processes. In this context, the computational technique employed is known as Brownian dynamics, which is, as its name suggests, based on the Brownian motion. This approach is particularly relevant for estimating the diffusive behaviour of molecules subjected to forces arising from intermolecular and surface interactions in biological systems, such as distinct protein mutants, see, e.g., Ref.\,\cite{biosystems}. Since Brownian motion is related to the irregular and fluctuating motion of a microscopic particle suspended in a fluid, caused by fluctuations of the surrounding molecules, it is natural to expect a peculiar distribution of velocities of the particles. In this context, the velocity autocorrelation function (VACF) quantifies how strongly the velocity of a particle in Brownian motion in a time $t$ remains correlated with its initial velocity. In other words, it quantifies the ``memory'' of the particle's motion.
In what follows we discuss its application to the VACF in Brownian motion based on the experimental data set reported in Ref.\,\cite{brownian1}. 

\section{Velocity autocorrelation function and hyperstatistics}\label{VACFS}
Following the applications of hyperstatistics reported in Ref.\,\cite{arxivhyper}, we explore here the motion of an optically trapped sphere confined near a wall. In such a case, the sphere's VACF exhibits a characteristic temporal decay. Such decay process depends on the sphere's proximity to the wall, since the hydrodynamic perturbation induced by the sphere becomes more significant depending on its position with respect to the wall. 
\begin{figure}[b!]
\centering
\includegraphics[width=0.85\textwidth]{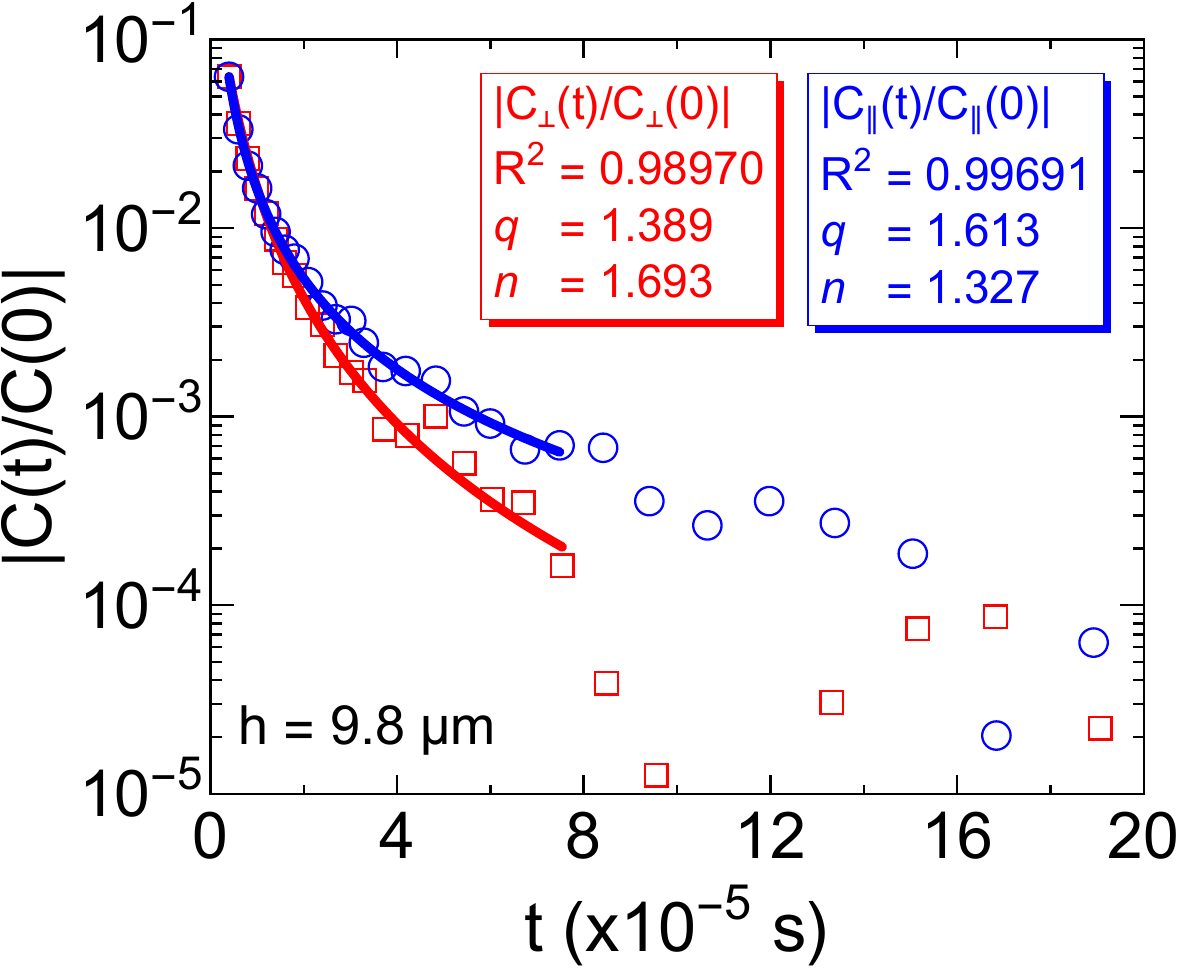}
\caption{\footnotesize \textbf{Velocity autocorrelation function in Brownian motion.} Normalized velocity autocorrelation function (VACF) as a function of time $t$, namely $|C_{\parallel}(t)/C_{\parallel}(0)|$ and $|C_{\perp}(t)/C_{\perp}(0)|$, considering a silica particle with radius $a_p = 1.5$\,$\mu$m placed at a distance $h = 9.8$\,$\mu$m of a silica sphere with radius $a_s = 100$\,$\mu$m. The experimental data set was extracted from Ref.\,\cite{brownian1} and fitted considering and adapted version of $B_q^n$, cf.\,Eq.\,\ref{BqE}, for both data sets \cite{arxivhyper}. The values of R$^2$, $q$, and $n$ are shown in the corresponding colors.}
\label{Fig-4}
\end{figure}
Following discussions in Refs.\,\cite{brownian1, brownian2, clercx1992, felderhof2005}, the decay of the VACF is dictated by a power law in time, which in turn depends on whether the disturbance produced by the sphere has already reached the wall and modified the fluid flow. 
Hence, for time scales smaller than the propagation time, hyperstatistics may be better suited for exploring VACF. The experimental data set in Fig.\,\ref{Fig-4}, taken from Ref.\,\cite{brownian1}, depicts the normalized VACF as a function of time regarding the motion of an optically trapped sphere constrained by the vicinity of a wall, cf.\,Fig.\,1 of Ref.\,\cite{brownian1}. The blue circles correspond to the VACF for the parallel component of the velocity, while the red squares correspond to the perpendicular one. Both experimental data sets were fitted employing hyperstatistics. It is to be noted that we have considered a time window smaller than the propagation time associated with the particle from its initial position to the wall. This is because for times larger than that, the hydrodynamic memory effects become significant \cite{brownian1}, as can be seen in Fig.\,\ref{Fig-4} for $t \gtrsim (8\times10^{-5})$\,s. The obtained values of $q$, namely 1.389 and 1.613, respectively for the perpendicular and parallel components of the velocity reveal that the VACF follows an intrinsic behaviour captured by hyperstatistics, being the decay of the parallel component slower than the perpendicular one.  
 A higher value of $q$ was obtained for the parallel component because, as the sphere approaches the wall, its motion generates a hydrodynamic perturbation that is reflected by the wall, thereby influencing the particle's movement. Hence, the higher value of $q$ for the parallel component indicates a greater complexity when compared with the perpendicular component, which ``feels less'' such a perturbation. This is also corroborated by the fact that the value of $n$ for the parallel component is lower, which indicates a more heterogeneous behaviour. This is because, in the analyzed regime, the perturbation in the VACF is sufficiently pronounced for it to deviate substantially from its average. Note that $n$ also plays the role of the ``weight'' on the distributed physical quantity, which can be visualized in the term $\exp(-n\beta_i/\langle \beta_{i} \rangle)$ in the $\gamma$-distribution function, cf.\,Eq.\,\ref{Laplacehyper}. Next, we discuss how hyperstatistics may provide additional insights into brain dynamics and the critical brain hypothesis (CBH).

\section{Brain dynamics and hyperstatistics}\label{Brain}
As it will be discussed below, the above interpretation naturally connects hyperstatistics with multilevel approaches in computational neuroscience and whole-brain emulation. In the classical Hodgkin-Huxley formalism, the elementary excitable unit is described in terms of membrane capacitance and voltage-dependent ionic conductances, particularly sodium and potassium currents, thereby providing a mechanistic account of the generation and propagation of action potentials \cite{Hodgkin-1952}. At a more computationally economical level, reduced spiking-neuron models, such as the Izhikevich model, reproduce a broad repertoire of cortical firing patterns, namely tonic spiking, bursting, spike-frequency adaptation, rebound responses, resonance, bistability, and inhibition-induced activity, while remaining suitable for large-scale network simulations, cf.\,Refs. \cite{Izhikevich-2003, Izhikevich-2004}. Whole-brain emulation frameworks extend this logic further by asking which level of biological detail is sufficient to preserve the relevant causal dynamics of brain function, ranging from ion-channel electrophysiology to spiking networks, synaptic adaptation, connectomics, population dynamics, and eventually whole-brain behaviour \cite{Sandberg-2008}.
From this perspective, hyperstatistics may be regarded as an emergent statistical descriptor of dynamics generated by lower-level neural models. Hodgkin-Huxley-type models describe how individual membranes become excitable; Izhikevich-type models preserve relevant spiking phenotypes with lower computational cost; connectome-based and whole-brain emulation models describe how such units interact through anatomical and functional connectivity; and hyperstatistics characterizes the resulting collective dynamics through $B_q^{n}$ at the domain level and $B_H$ for the macroscopic one \cite{arxivhyper}, avalanche-size distributions, and deviations from ordinary Boltzmann-Gibbs statistics at the domain level. Thus, both $q$ and $n$ parameters do not need to encode a molecular, cellular, or structural property directly. Rather, such parameters may quantify how the integrated system behaves once microscopic neuronal events are coupled across mesoscopic domains.
This also emphasizes an important independence of the present approach from any unique molecular or structural interpretation of brain function. Note that hyperstatistics does not require that the relevant physical domains correspond specifically to ion channels, synapses, minicolumns, cortical areas, or connectomic modules. These may all represent valid levels of biological description, but the hyperstatistical treatment is concerned with the distribution of effective dynamical domains and their collective statistical signature. In this sense, the same $B_q^n$ could in principle emerge from different microscopic implementations, provided that they preserve the relevant propagation, memory, coupling, and fluctuation structure of the system. This is compatible with the scale-separation problem discussed in whole-brain emulation: if the relevant causal dynamics are preserved above a given resolution, then the macroscopic critical-like signatures should be reproducible without explicitly simulating all lower molecular degrees of freedom \cite{Sandberg-2008}. Such a multilevel interpretation suggests a possible clinical and functional extension. If the healthy conscious brain operates near a critical regime, then pathological or pharmacologically altered brain states may be interpreted as departures from this critical window. A subcritical-like state would correspond to insufficient propagation of neuronal activity, reduced avalanche sizes, impaired long-range correlations, and diminished information integration. This picture is compatible with observations that unconsciousness and some anesthetic states are associated with departure from critical dynamics and restriction of avalanche size and duration, see, e.g., Refs.\ \cite{Tagliazucchi-2016,Varley-2020}. In this framework, coma or deep disorders of consciousness may be viewed, at least phenomenologically, as subcritical-like dynamical regimes, in which neuronal cascades fail to propagate efficiently through the large-scale network.
Conversely, supercritical-like regimes may correspond to excessive propagation, unstable amplification, and reduced control of associative transitions. Speculatively, psychosis could be approached from this standpoint as a state in which internally generated activity spreads too broadly or too persistently across associative networks, favoring disorganized thought, aberrant salience, hallucinations, or delusional binding. Such interpretation remains hypothetical, but it is consistent with the broader literature suggesting altered criticality in schizophrenia \cite{Alamian-2022}. Certain drug-induced altered states may also be interpreted along this axis. Psychedelic compounds and ketamine have been associated with increased neural signal diversity and altered large-scale brain dynamics \cite{Schartner-2017, Shew-2009}, suggesting that pharmacological modulation of excitation, inhibition, neuromodulation, and network coupling may shift the brain away from ordinary waking criticality.
Therefore, hyperstatistics may offer a compact bridge between microscopic biophysics, computational emulation, and clinical brain states. At one extreme, detailed Hodgkin-Huxley-type models provide mechanistic grounding at the membrane level; at another, whole-brain emulation seeks to reproduce sufficient causal structure for global brain dynamics; between them, reduced spiking models provide scalable approximations of neuronal behaviour. Hyperstatistics enters at the emergent level, asking whether the resulting system displays $B_q^{n}$-type behavior at the domain level, i.e, specific regions of the brain; long-range correlations, and critical-like avalanche behaviour. Under this view, $q$ and $n$ could potentially serve as key indexes of the brain's dynamical regime, with values near the critical maximum reflecting optimal complexity, whereas deviations toward lower or altered $(q, n)$-patterns may indicate subcritical or supercritical pathological states, see Fig.\,\ref{Fig-1.1}. Future work should therefore compare systematically $q$ and $n$, cf.\,Table \ref{Table-2}, across simulated neural systems, empirical EEG/MEG/fMRI recordings, anesthesia, coma, schizophrenia, psychedelic states, and other neuropsychiatric conditions using standardized avalanche-detection and fitting protocols.

Recent studies by Korchinski and collaborators \cite{korchinski2021} indicate that brain activity occurs by means of avalanches of neuronal activation in the form of cascades, i.e., flows of neuronal activations that spread through the inherent network during successive time steps. The size $s$ of an avalanche corresponds to the total number of activated neurons involved in that cascade, over which the cascade remains active. The authors of Ref.\,\cite{korchinski2021} consider that the probability distribution $P(s)$ describes how likely avalanches of different sizes are, so that for each $s$ there is a corresponding spreading time. Avalanches with the same size may last for different times, depending on how the activity propagates through the complex network inherent to the brain's dynamics, indicating thus the need for a theoretical framework that takes into account a distribution of distinct characteristic times for each avalanche. Precisely because the consideration of homogeneous conductances and single response time scales breaks down when an avalanche spreads through a disordered network, a description in terms of hyperstatistics \cite{arxivhyper} consists an appropriate framework when membrane responses involve heterogeneous environments and distributed relaxation processes as it occurs in the brain.

\subsection{Hyperstatistics and possible brain criticality}\label{cbh}
\begin{figure}
\centering
\includegraphics[width=0.82\textwidth]{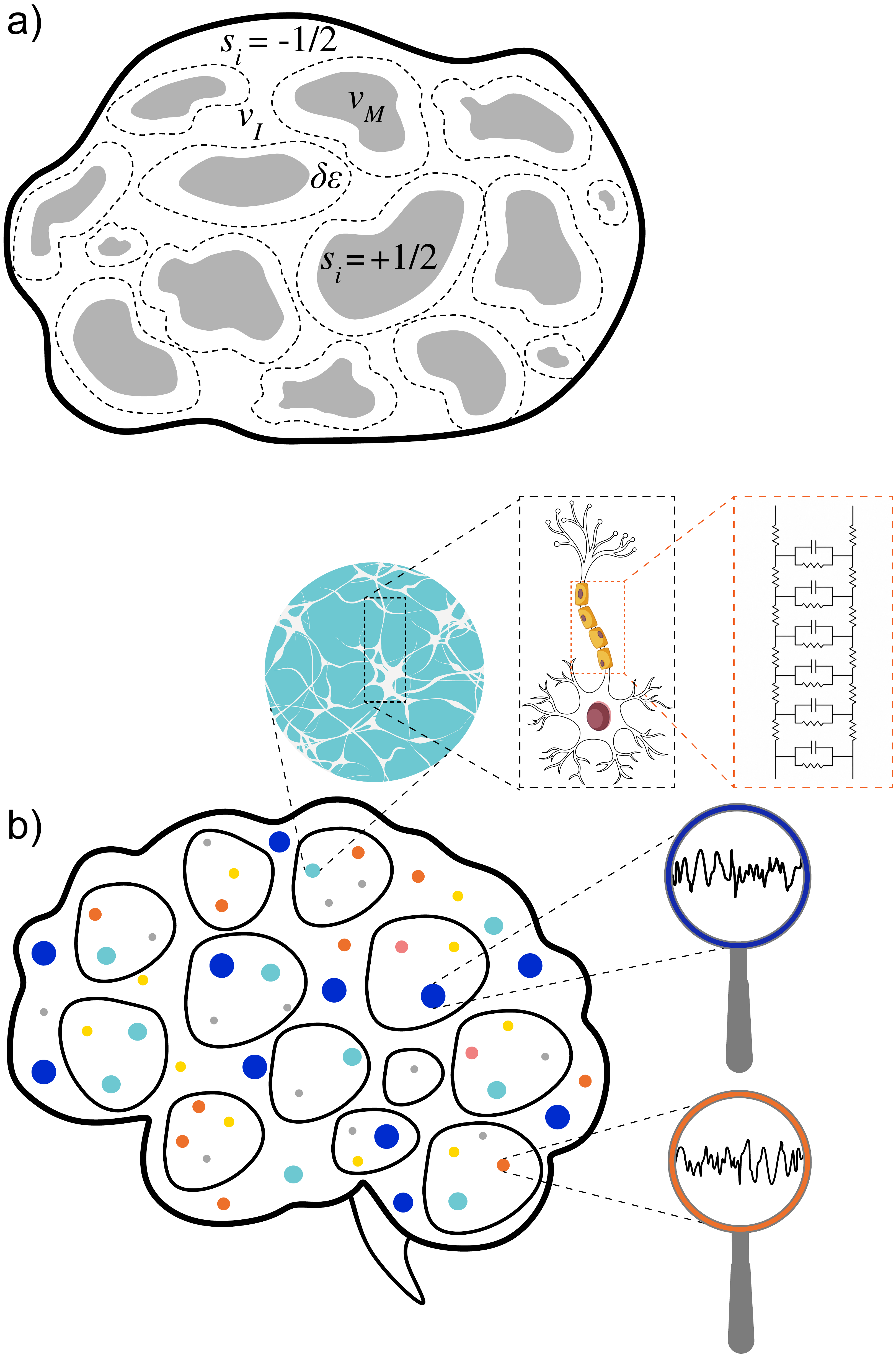}
\caption{\footnotesize \textbf{The Mott transition, brain dynamics, and hyperstatistics.} a) Schematic representation of the phases-coexistence region of the Mott transition \cite{GR1}. The gray regions correspond to metallic domains with volume $v_M$ associated with spin $s_i = +1/2$, which are embedded in a Mott-insulating matrix of volume $v_I$ associated with $s_i = -1/2$. The interaction between neighbouring metallic domains $\delta\varepsilon$ is also indicated. Figure extracted from Ref.\,\cite{GR1} aiming to illustrate the similarity between the Griffiths phase in condensed matter Physics and in the brain. b) Schematic representation of a brain composed of domains with distinct neuronal activities indicated by the nodes in different colors. Each domain contains a distribution of neuronal connectivities, giving rise to distinct characteristic neuronal signals, represented by the magnifiers with different colored frames. Such a representation captures the microscopic, mesoscopic, and macroscopic scales of the system \cite{arxivhyper}, highlighting that hyperstatistics provides a natural and appropriate hierarchical framework for describing the complexity of brain dynamics. A magnified view of one of the nodes depicts a neuronal network in which a single neuron is highlighted. Its axonal membrane is further represented by an equivalent electrical circuit containing resistive and capacitive elements, cf.\,Hodgkin-Huxley model \cite{Hodgkin-1952}.} 
\label{Fig-5}
\end{figure}
It has been discussed in the literature that a Griffiths phase may be closely linked to the CBH \cite{munoz}. In this scenario, the brain embodies rare regions, i.e., domains, with slightly distinct neuronal activities, cf. Fig.\,\ref{Fig-5}. Hence, as pointed out previously, hyperstatistics can be seen as an appropriate framework to describe the complex behaviour of neuronal activity, where heterogeneity and structural disorder coexist. Following discussions in Ref.\,\cite{munoz}, the Griffiths phase in the context of brain dynamics is characterized mainly by two factors: heterogeneity and structural disorder. In condensed matter Physics, the fingerprints of the Griffiths phase are: random spatial disorder and long relaxation times \cite{Vojta2006,GR1}. Such a phase is composed of the so-called rare regions (or domains) embedded in a matrix, e.g., rare magnetized regions within a non-magnetized matrix for the magnetic Griffiths phase \cite{Griffiths,Vojta2006}. At the onset of a Griffiths phase, there is an inherent slowing down of the system \cite{cell,GR1}, which can be computed in terms of the so-called autocorrelation function $C(t)$ \cite{Vojta2006}. The latter decays more slowly with time $t$ in the Griffiths region than in the non-Griffiths regime \cite{Vojta2006}. In simple terms, $C(t)$ quantifies the degree to which a variable at $t = 0$ remains correlated with itself after a time interval $t$, being the value of the spin projection the variable of interest when analyzing the magnetic Griffiths phase \cite{Vojta2006}. The rare region part of $C(t)$, i.e., $C_{RR}(t)$, can be computed employing \cite{Vojta2006}:
\begin{equation}
C_{RR}(t) \sim \int_0^{\infty} \exp[-t/\xi_t(L_{RR},r_0)]  w(L_{RR}) dL_{RR},
\label{crr}
\end{equation}
where $L_{RR}$ is the size of a rare region, $w(L_{RR})$ the probability of finding a rare region,  $\xi_t$ the relaxation time, and $r_0$ a reduced temperature. In the frame of the diluted Ising model used to describe a magnetic Griffiths phase and following discussions in Ref.\,\cite{Vojta2006}, the probability of finding a rare region of linear size \(L_{RR}\) is given by $ w(L_{RR})=\exp\left(-\tilde{p}L_{RR}^{d}\right)$, where $\tilde{p}=-\ln(1-p)$, being $p$ the probability of finding a vacancy at a given site of the rare region, and $d$ the dimensionality of the system. Remarkably, Eq.\,\ref{crr} has the same mathematical structure of the Laplace transform of the $\gamma$-distribution, cf.\,Eq.\,\ref{laplace2}, being that the rare regions play the role of the domains, cf.\,Figs.\,\ref{Fig-5} a) and b). Figure\,\ref{Fig-5} a) depicts metallic puddles embedded in an insulating matrix in the onset of the Mott transition, cf.\,inset of Fig.\,1 in Ref.\,\cite{GR1}, which is analogous to domains in the brain with distinct neuronal activities, cf.\,Fig.\,\ref{Fig-5} b). Equation\,\ref{crr} makes it evident that a distribution of $L_{RR}$ is considered, which in turn can be associated with a distribution of $\xi_t(L_{RR})$. This is precisely the central idea underlying Eq.\,\ref{laplace2}, which describes a distribution of relaxation times, for instance, in the context of a capacitor discharge \cite{arxivhyper}. Next, we discuss the connection between Griffiths phases, CBH, and hyperstatistics. We recall that in the context of brain dynamics, neuronal avalanches correspond to cascades of brain activity propagating through networks and are associated with the CBH \cite{korchinski2021} and optimal information processing \cite{kinouchi2006, dahmen2019}. Under this perspective, the brain is proposed to operate near a critical point, analogously to critical endpoints in condensed matter Physics, see, e.g., Refs.\,\cite{breakdownscaling,mrb,prbl2025}, separating ordered and disordered dynamical regimes, where information processing is optimized \cite{kinouchi2006}. The authors of Ref.\,\cite{munoz} discuss the possibility of a Griffiths phase in neural networks, demonstrating with numerical calculations that, right at the onset of such a phase, the dynamical susceptibility is dramatically enhanced and the system is \emph{anomalously slow}. In Ref.\,\cite{munoz}, the authors estimate the slow-dynamics through a saddle-point approximation. The hyperstatistics approach is distinct in the sense that it considers a distribution of relaxation times, so that the system can be described by $B_q^n$ at the domain level, which in turn incorporates $\langle \tau \rangle$, $q$ and $n$, cf.\,discussions in Section\,\ref{qandn}. In this context, hyperstatistics may provide a more accurate description of the broad distribution of relaxation times associated with rare-region dynamics in the Griffiths phase. In the following, we make use of the numerical data set of a distribution of avalanche sizes reported in Ref.\,\cite{korchinski2021} to showcase our approach. 
\begin{figure}[t!]
\centering
\includegraphics[width=0.53\textwidth]{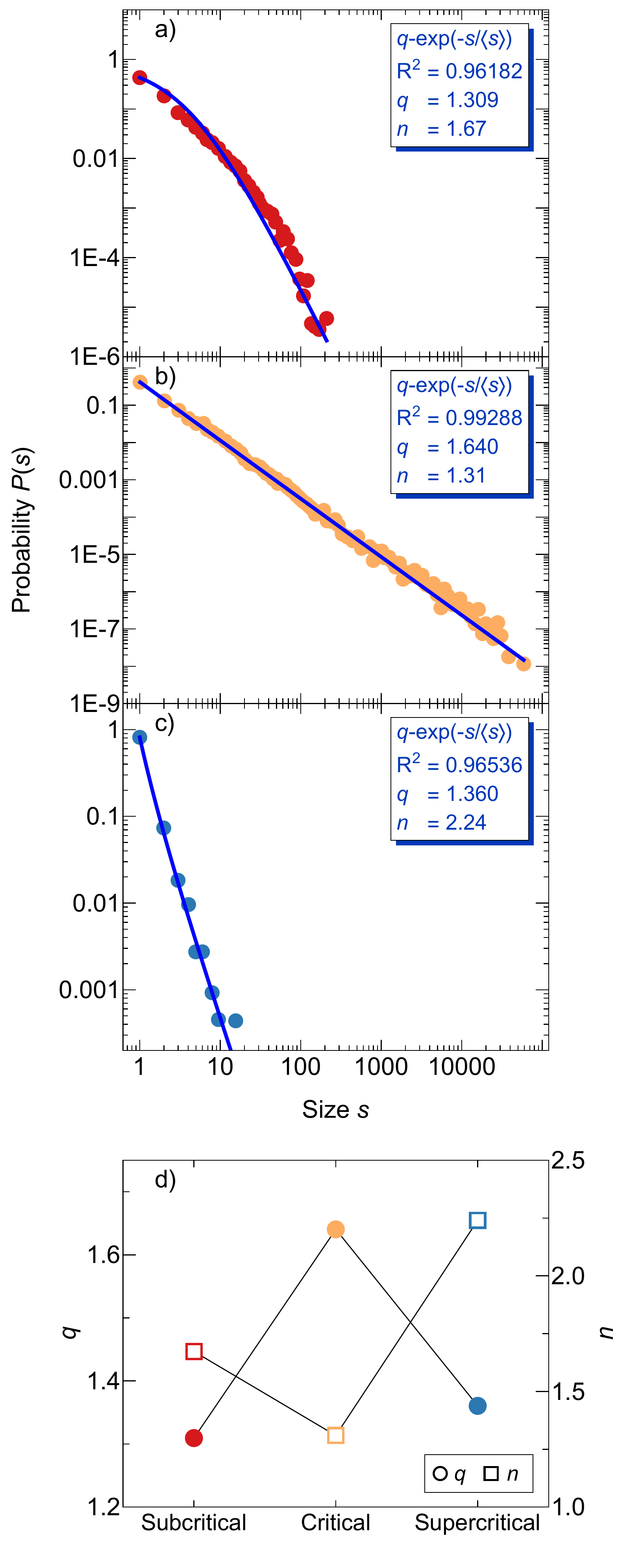}
\caption{\footnotesize \textbf{Probability of observing an avalanche with size $s$ as a function of $s$.} a) subcritical, b) critical, and c) supercritical regimes considering a network of size $N = 10^5$. The data set was extracted from Ref.\,\cite{korchinski2021} and fitted employing the hyperstatistics approach \cite{arxivhyper}. The obtained values of R$^2$, $q$, and $n$ are depicted in the corresponding colors of the analyzed data set for each case. d) Obtained values of $q$ (circles, left axis) and $n$ (squares, right axis) for the subcritical, critical and supercritical regimes.}
\label{Fig-6}
\end{figure} 
Essentially, in Ref.\,\cite{korchinski2021}, the authors have explored the transition between subcritical, critical, and supercritical regimes. In the subcritical regime, avalanches are comparatively small and the probability of percolation goes fast to zero as the avalanche size is increased.  In the critical regime in turn, avalanches follow a power-law-type decay, whereas in the supercritical regime avalanches spread through a significant portion of the network. Hence, given its inherent pronounced complexity, brain dynamics can be seen as an appropriate physical context for the application of hyperstatistics.  Figure\,\ref{Fig-6} depicts the probability of observing an avalanche with size $s$, i.e., the probability that a spreading cascade contains $s$ activated nodes, as a function of $s$. Note that a distribution of propagation time is associated with each avalanche size. According to Ref.\,\cite{korchinski2021}, the subcritical regime is characterized by finite-size avalanches, while the supercritical regime by the emergence of permanent giant avalanches with a large characteristic scale. However, right at the critical regime, scale-free avalanches appear, which is captured by the minimum value of $n$, i.e., $n = 1.31$ and the enhanced value of $q = 1.640$. The obtained values of $q$ and $n$ indicate that the system is most heterogeneous right at the critical regime, as reflected by the minimum of $n$ and the maximum of $q$, cf.\,Fig.\,\ref{Fig-6} d), which is expected at critical points.
Interestingly, since in the asymptotic regime the $q$-exp function reduces to a power law, namely $\exp_q(-x) \sim x^{-1/(q-1)}$ \cite{arxivhyper}, the value of $q$ obtained at the critical point for the numerical data set extracted from Ref.\,\cite{korchinski2021} yields an exponent $1/(q-1) \simeq 1.56$, comparable to the asymptotic power-law exponent $3/2$, i.e., $P(s) \sim s^{-3/2}$, reported for the data set of Fig.\,2 of Ref.\,\cite{korchinski2021}. Hence, our analysis points towards the CBH, given that the value of $q$ ($n$) is maximized (minimized) precisely at the critical point. It is worth mentioning that, in the present work, we restrict our analysis to numerical data due to the inherent difficulties associated with performing such delicate experiments, which often lead to non-reproducible results. Hence, it is highly desirable to establish a systematic set of experimental, i.e., clinical data, based on a well-defined protocol to investigate both the vicinity and the critical point, aiming to probe the CBH.

\section{Conclusions and perspectives}
The hyperstatistics approach is a novel statistical mechanics framework for describing complex systems. Some additional remarks were made on hyperstatistics regarding its original proposal in Ref.\,\cite{arxivhyper}, being its application extended to the velocity autocorrelation function in the frame of Brownian motion and avalanches in brain networks regarding the critical brain hypothesis. Also, we have discussed its potential to describe brain dynamics in terms of states of consciousness, which could also impact the interpretation of diagnostic test results regarding, for instance, neuropsychiatric conditions. We also briefly discussed how hyperstatistics could provide a suitable framework for properly describing brain criticality. Many other phenomena, including additional high-energy collisions at LHC, apart from that explored in Ref.\,\cite{arxivhyper}, particle movement in fluids under extreme conditions, photoluminescence decay \cite{PL}, among others, are on the agenda and constitute part of ongoing projects.

\section*{Acknowledgements}
MdeS acknowledges partial financial support from the S\~ao Paulo Research Foundation -- Fapesp (Grants 2011/22050-4, 2017/07845-7, and 2019/24696-0), National Council for Scientific and Technological Development -- CNPq (Grant 303772/2023-9), and Capes -- Finance Code 001 (M.\,Sc.\,fellowship of SMS). GL acknowledges financial support from the S\~ao Paulo Research Foundation -- Fapesp (Grant 2018/18900-1).
MdeS, SMS, and LS thank Prof. C. Tsallis for mentioning the terminology ``micro, meso, and macro'', employed in Table \ref{Table-1}, during a discussion about hyperstatistics in his office at CBPF-RJ on April 13th 2026.
\section*{Data availability}
The data that support the findings of this work are openly available \cite{zenodo}.

\section*{Author contributions}
MdeS carried out the calculations with inputs from SMS and LS. SMS and LS carried out the data fits. GL wrote Section\,\ref{Brain}. MdeS wrote the paper with contributions from SMS, LS, and GL. All authors revised the manuscript. MdeS conceived and supervised the whole project.


\begin{thebibliography}{00}

\bibitem{reviewmariano} R.S. Manna, B. Wolf, M. de Souza, M. Lang, High-resolution thermal expansion measurements under helium-gas
pressure,  Rev. Sci. Instrum. \textbf{83}, 085111 (2012). http://dx.doi.org/10.1063/1.4747272

\bibitem{valvano1992} J.W. Valvano, Temperature measurements, Adv. Heat Transf. \textbf{22}, 359 (1992). https://doi.org/10.1016/S0065-2717(08)70346-0 

\bibitem{hitosugi2016} Y. Okada, Y. Ando, R. Shimizu, E. Minamitani, S. Shiraki, S. Watanabe, T. Hitosugi, Scanning tunnelling spectroscopy of
superconductivity on surfaces of LiTi$_2$O$_4$(111) thin films, Nat. Commun. \textbf{8}, 15975 (2017). https://doi.org/10.1038/ncomms15975

\bibitem{prl07}
M. de Souza, A. Br\"uhl, Ch. Strack, B. Wolf, D. Schweitzer, M. Lang, 
Anomalous Lattice Response at the Mott Transition in a Quasi-2D Organic Conductor, Phys. Rev. Lett. \textbf{99}, 037003 (2007).
https://doi.org/10.1103/PhysRevLett.99.037003

\bibitem{GR1} I.F. Mello, L. Squillante, G.O. Gomes, A.C. Seridonio, M. de Souza, Griffiths-like phase close to the Mott transition, J. Appl. Phys. \textbf{128}, 225102 (2020).
https://doi.org/10.1063/5.0018604

\bibitem{arxivhyper} L. Squillante, S.M. Soares, C. Tsallis, M. de Souza, Hyperstatistics, arXiv:2604.24783v5 (2026).
https://doi.org/10.48550/arXiv.2604.24783

\bibitem{wilk2000}
G. Wilk, Z. W{\l}odarczyk, Interpretation of the Nonextensivity Parameter $q$ in Some Applications of Tsallis Statistics and L\'evy Distributions, Phys. Rev. Lett. \textbf{84}, 2770 (2000).
https://doi.org/10.1103/PhysRevLett.84.2770 

 \bibitem{ricieribook} A.P. Ricieri, C\'alculo fracion\'ario, transformada de Laplace e outros bichos, Prandiano - Matem\'atica aplicada \`a vida (1993).
 
 \bibitem{tsallis} C. Tsallis, Possible generalization of Boltzmann-Gibbs statistics, J. Stat. Phys. \textbf{52}, 479 (1988). https://doi.org/10.1007/BF01016429

\bibitem{boltzmann} L. Boltzmann, On the relationship between the second fundamental theorem of the mechanical theory of heat and probability calculations regarding the conditions for thermal equilibrium, Kaiserlichen Akademie der Wissenschaften \textbf{76}, 373 (1877).

\bibitem{gibbs} J.W. Gibbs, On the fundamental formula of statistical mechanics, with applications to astronomy and thermodynamics, Proceedings of the American Association for the Advancement of Science \textbf{33}, 57 (1884).

 \bibitem{beck2003} C. Beck, E.G.D. Cohen, Superstatistics, Physica A \textbf{322}, 267 (2003). https://doi.org/10.1016/S0378-4371(03)00019-0
     
\bibitem{descartes} R. Descartes, Discours de la m\'ethode pour bien conduire sa raison, et chercher la v\'erit\'e dans les sciences, Jan Maire, (1637).

\bibitem{boumali2026} A. Boumali, Hyperstatistical thermodynamics of the one-dimensional Klein-Gordon and Dirac oscillators: a closed-form $q$-generalized Boltzmann factor and a quantitative comparison with Beck's superstatistics, arXiv:2606.12454 (2026). 
https://doi.org/10.48550/arXiv.2606.12454

\bibitem{nicola} N. Fabiano, Non-Markovian Quantum Decay in Complex Environments: A Hyperstatistical Approach, arXiv:2609.22190 (2026). 
https://doi.org/10.48550/arXiv.2609.22190

\bibitem{aboumalilandau} A. Boumali, Third-cumulant signatures of hyperstatistical temperature fluctuations in collisionless-plasma Landau damping, Physica A \textbf{702}, 132088 (2026). https://doi.org/10.1016/j.physa.2026.132088

\bibitem{alice} B. Abelev \emph{et al.} (ALICE Collaboration), Transverse momentum distribution and nuclear modification factor of charged particles in $p$-Pb collisions at $\sqrt{s_{NN}}$ = 5.02\,TeV, Phys. Rev. Lett. \textbf{110}, 082302 (2013).
https://doi.org/10.1103/PhysRevLett.110.082302

\bibitem{bodenschatz} A. La Porta, G.A. Voth, A.M. Crawford, J. Alexander, E. Bodenschatz, Fluid particle accelerations in fully developed turbulence, Nature \textbf{409}, 1017 (2001). https://doi.org/10.1038/35059027

\bibitem{brownian1}
S. Jeney, B. Luki{\'c}, J.A. Kraus, T. Franosch, L. Forr{\'o}, Anisotropic Memory Effects in Confined Colloidal Diffusion, Phys. Rev. Lett. \textbf{100}, 240604 (2008).
https://doi.org/10.1103/PhysRevLett.100.240604

\bibitem{youtubevideo} Conhecimentos de F\'{\i}sica (CdF) - \textit{Hyperstatistics} - \url{https://www.youtube.com/watch?v=WkRCCGHI9Ws}.

\bibitem{korchinski2021} D.J. Korchinski, J.G. Orlandi, S.-W. Son, J. Davidsen, Criticality in Spreading Processes without Timescale Separation and the Critical Brain Hypothesis, Phys. Rev. X \textbf{11}, 021059 (2021). https://doi.org/10.1103/PhysRevX.11.021059

\bibitem{BJP} M. de Souza, R. Paupitz, A. Seridonio, R.E. Lagos, Specific Heat Anomalies in Solids Described by a Multilevel Model, Braz. J. Phys. \textbf{46}, 206 (2016). 
https://doi.org/10.1007/s13538-016-0404-9

\bibitem{Jin} H. Jin, R.R. Ullah, P. Klavins, V. Taufour, Importance of anisotropic interactions for hard-axis or hard-plane ordering of Ce-based ferromagnets. Phys. Rev. B \textbf{112}, 024437 (2025).
https://doi.org/10.1103/99ww-msjh
 
 \bibitem{biosystems}
A. Mu\~niz-Chicharro, L.W. Votapka, R.E. Amaro, R.C. Wade, Brownian dynamics simulations of biomolecular diffusional association processes, WIREs Comput. Mol. Sci. \textbf{13}, e1649 (2023). https://doi.org/10.1002/wcms.1649 

\bibitem{brownian2}
K. Huang, I. Szlufarska, Effect of interfaces on the nearby Brownian motion, Nat. Commun. \textbf{6}, 8558 (2015).
https://doi.org/10.1038/ncomms9558

\bibitem{clercx1992}
H.J.H. Clercx, P.P.J.M. Schram, Brownian particles in shear flow and harmonic potentials: A study of long-time tails, Phys. Rev. A \textbf{46}, 1942 (1992).
https://doi.org/10.1103/PhysRevA.46.1942

\bibitem{felderhof2005}
B.U. Felderhof, Effect of the Wall on the Velocity Autocorrelation Function and Long-Time Tail of Brownian Motion, J. Phys. Chem. B \textbf{109}, 21406 (2005).
https://doi.org/10.1021/jp051335b 

\bibitem{Hodgkin-1952} A.L. Hodgkin, A.F. Huxley, A quantitative description of membrane current and its application to conduction and excitation in nerve, J. Physiol. \textbf{117}, 500 (1952). https://doi.org/10.1113/jphysiol.1952.sp004764

\bibitem{Izhikevich-2003} E.M. Izhikevich, Simple model of spiking neurons, IEEE Trans. Neural Netw. \textbf{14}, 1569 (2003). https://doi.org/10.1109/TNN.2003.820440

\bibitem{Izhikevich-2004} E.M. Izhikevich, Which model to use for cortical spiking neurons?, IEEE Trans. Neural Netw. \textbf{15}, 1063 (2004). https://doi.org/10.1109/TNN.2004.832719

\bibitem{Sandberg-2008} A. Sandberg, N. Bostrom, Whole Brain Emulation: A Roadmap, Future of Humanity Institute, Oxford University, Oxford (2008).

\bibitem{Tagliazucchi-2016} E. Tagliazucchi, D.R. Chialvo, M. Siniatchkin, E. Amico, J.F. Brichant, V. Bonhomme \textit{et al.}, Large-scale signatures of unconsciousness are consistent with a departure from critical dynamics, J. R. Soc. Interface \textbf{13}, 20151027 (2016). https://doi.org/10.1098/rsif.2015.1027 

\bibitem{Varley-2020} T.F. Varley, O. Sporns, A. Puce, J. Beggs, Differential effects of propofol and ketamine on critical brain dynamics, PLoS Comput. Biol. \textbf{16}, e1008418 (2020). https://doi.org/10.1371/journal.pcbi.1008418

\bibitem{Alamian-2022} G. Alamian, T. Lajnef, A. Pascarella, J.M. Lina, L. Knight, K. Jerbi \textit{et al.}, Altered brain criticality in schizophrenia: new insights from magnetoencephalography, Front. Neural Circuits \textbf{16}, 630621 (2022). https://doi.org/10.3389/fncir.2022.630621

\bibitem{Schartner-2017} M.M. Schartner, R.L. Carhart-Harris, A.B. Barrett, A.K. Seth, S.D. Muthukumaraswamy, Increased spontaneous MEG signal diversity for psychoactive doses of ketamine, LSD and psilocybin, Sci. Rep. \textbf{7}, 46421 (2017). https://doi.org/10.1038/srep46421

\bibitem{Shew-2009} W.L. Shew, H. Yang, T. Petermann, R. Roy, D. Plenz, Neuronal avalanches imply maximum dynamic range in cortical networks at criticality, J. Neurosci. \textbf{29}, 15595 (2009). https://doi.org/10.1523/jneurosci.3864-09.2009 

\bibitem{munoz} P. Moretti, M.A. Mu\~noz, Griffiths phases and the stretching of criticality in brain networks, Nat. Commun. \textbf{4}, 2521 (2013). https://doi.org/10.1038/ncomms3521

\bibitem{Vojta2006} T. Vojta, Rare region effects at classical, quantum and nonequilibrium phase transitions, J. Phys. A: Math. Gen. \textbf{39}, R143 (2006). https://doi.org/10.1088/0305-4470/39/22/R01


\bibitem{Griffiths} R.B. Griffiths, Nonanalytic behavior above the critical point in a random Ising ferromagnet, Phys. Rev. Lett. \textbf{23}, 17 (1969). https://doi.org/10.1103/PhysRevLett.23.17

\bibitem{cell} L. Squillante, I.F. Mello, L.S. Ricco, M.F. Minicucci, A.M. Ukpong, A.C. Seridonio, R.E. Lagos-Monaco, M. de Souza, Cellular Griffiths-like phase, Heliyon \textbf{10}, e34622 (2024). https://doi.org/10.1016/j.heliyon.2024.e34622


\bibitem{kinouchi2006} O. Kinouchi, M. Copelli, Optimal Dynamical Range of Excitable Networks at Criticality, Nat. Phys. \textbf{2}, 348 (2006). https://doi.org/10.1038/nphys289

\bibitem{dahmen2019} D. Dahmen, S. Gr\"{u}n, M. Diesmann, M. Helias, Second Type of Criticality in the Brain Uncovers Rich Multiple-Neuron Dynamics, Proc. Natl. Acad. Sci. U.S.A. \textbf{116}, 13051 (2019). https://doi.org/10.1073/pnas.1818972116

\bibitem{breakdownscaling} L. Bartosch, M. de Souza, M. Lang, Scaling theory of the Mott transition and breakdown of the Gr\"uneisen scaling near a finite-temperature critical end point, Phys. Rev. Lett. \textbf{104}, 245701 (2010). https://doi.org/10.1103/PhysRevLett.104.245701

\bibitem{mrb} L. Squillante, I.F. Mello, A.C. Seridonio, M. de Souza, Giant caloric effects close to any critical end point, Mat. Res. Bull. \textbf{142}, 111413 (2021). https://doi.org/10.1016/j.materresbull.2021.111413

\bibitem{prbl2025} S.M. Soares, L. Squillante, H.S. Lima, C. Tsallis, M. de Souza, Universally nondiverging Gr\"uneisen parameter at critical points, Phys. Rev. B \textbf{111}, L060409 (2025). https://doi.org/10.1103/PhysRevB.111.L060409


\bibitem{PL}
J.R. Martins, V. Krivenkov, C.R. Bernardo, P. Samokhvalov, I. Nabiev, Y.P. Rakovich, M.I. Vasilevskiy, Statistical Analysis of Photoluminescence Decay Kinetics in Quantum Dot Ensembles: Effects of Inorganic Shell Composition and Environment,
J. Phys. Chem. C  \textbf{126}, 20480 (2022).
https://doi.org/10.1021/acs.jpcc.2c06134

\bibitem{zenodo} L. Squillante, S.M. Soares, G. Lepski, M. de Souza, A hierarchical statistical mechanics approach: hyperstatistics and its applications. Zenodo https://doi.org/10.5281/zenodo.20526858.

\end{thebibliography}
\end{document}